\documentclass[acmlarge]{acmart}

\makeatletter
\newcommand{\myconfshort}{\acmConference@shortname}
\newcommand{\myconffull}{\acmConference@name}
\newcommand{\myconfdate}{\acmConference@date}
\newcommand{\myconfloc}{\acmConference@venue}
\AtBeginDocument{
  \fancypagestyle{firstpagestyle}{
    \fancyhead{}%
    \fancyfoot[C]{}%
  }
  \fancyhf{}
  \fancyhead[LO]{\@headfootfont\shorttitle}%
  \fancyhead[RE]{\@headfootfont\@shortauthors}%
  \fancyhead[LE]{\@headfootfont\footnotesize \myconfshort, \myconfdate, \myconfloc}%
  \fancyhead[RO]{\@headfootfont\footnotesize \myconfshort, \myconfdate, \myconfloc}%
  \fancyfoot[C]{}%
}
\makeatother
\acmBooktitle{\conffull\@ (\confshort), \confdate, \confloc}

\AtBeginDocument{%
  }

\copyrightyear{2026}
\acmYear{2026}
\setcopyright{cc}
\setcctype{by}
\acmConference[FAccT '26]{The 2026 ACM Conference on Fairness, Accountability, and Transparency}{June 25--28, 2026}{Montreal, QC, Canada}
\acmBooktitle{The 2026 ACM Conference on Fairness, Accountability, and Transparency (FAccT '26), June 25--28, 2026, Montreal, QC, Canada}
\acmDOI{10.1145/3805689.3812289}
\acmISBN{979-8-4007-2596-8/2026/06}

\usepackage{xcolor}
\usepackage[normalem]{ulem}  
\newcommand{\revmode}{clean}  

\usepackage{ifthen}

\ifthenelse{\equal{\revmode}{clean}}{%
  \newcommand{\rev}[1]{#1}%
  \newcommand{\new}[1]{#1}%
  \newcommand{\removed}[1]{}%
  \newcommand{\removedm}[1]{}%
  \newenvironment{revblock}{}{}%
}{}

\ifthenelse{\equal{\revmode}{revision}}{%
  \newcommand{\rev}[1]{\textcolor{blue}{#1}}%
  \newcommand{\new}[1]{\textcolor{blue}{#1}}%
  \newcommand{\removed}[1]{}%
  \newcommand{\removedm}[1]{}%
  \newenvironment{revblock}{%
    \par\noindent\hspace{-0.5em}%
    \begin{minipage}[t]{2pt}\color{blue}\vrule width 2pt\end{minipage}%
    \hspace{0.5em}\begin{minipage}[t]{\dimexpr\linewidth-1.5em}%
  }{\end{minipage}\par}%
}{}

\ifthenelse{\equal{\revmode}{trackchanges}}{%
  \newcommand{\rev}[1]{\textcolor{blue}{#1}}%
  \newcommand{\new}[1]{\textcolor{blue}{#1}}%
  \newcommand{\removed}[1]{{\color{red}\sout{#1}}}%
  \newcommand{\removedm}[1]{\textcolor{red}{\sout{#1}}}%
  \newenvironment{revblock}{%
    \par\noindent\hspace{-0.5em}%
    \begin{minipage}[t]{2pt}\color{blue}\vrule width 2pt\end{minipage}%
    \hspace{0.5em}\begin{minipage}[t]{\dimexpr\linewidth-1.5em}%
  }{\end{minipage}\par}%
}{}

\begin{document}

\title[The Accountability Paradox]{The Accountability Paradox: How Platform API Restrictions Undermine AI Transparency Mandates}
\date{January 12, 2026}


\author{Florian A. D. Burnat}
\affiliation{%
  \institution{University of Bath}
  \city{Bath}
  \country{England, UK}}
\email{fadb20@bath.ac.uk}

\author{Brittany I. Davidson}
\affiliation{%
  \institution{University of Bath}
  \city{Bath}
  \country{England, UK}}
\email{bid23@bath.ac.uk}

\renewcommand{\shortauthors}{Burnat and Davidson}

\begin{abstract}
    Recent application programming interface (API) restrictions on major social media platforms challenge compliance with the EU Digital Services Act~\citep{European-Parliament-and-Council2022-wf}, which mandates data access for algorithmic transparency. We develop a structured audit framework to assess the growing misalignment between regulatory requirements and platform implementation. Our comparative analysis of X/Twitter, Reddit, TikTok, and Meta identifies critical ``audit blind-spots'' where platform content moderation and algorithmic amplification remain inaccessible to independent verification.

    We extend our analysis to examine all Very Large Online Platforms (VLOPs) designated under the DSA and compare regulatory approaches across jurisdictions, including the EU AI Act and the UK Online Safety Act. Through systematic documentary analysis of platform policies, API documentation, regulatory filings, and transparency reports, triangulated with published research on researcher access experiences, our findings reveal an ``accountability paradox'': as platforms increasingly rely on AI systems, they simultaneously restrict the capacity for independent oversight.

    We propose concrete technical and policy interventions aligned with the AI Risk Management Framework of the National Institute of Standards and Technology~\citep{NIST2023}, including privacy-preserving access mechanisms using differential privacy and secure enclaves, federated access models with graduated trust levels, and enhanced regulatory enforcement mechanisms with specific compliance metrics.
\end{abstract}





\maketitle

\section{Introduction}

Social media platforms have become integral to public discourse and are critical sources of training data for contemporary artificial intelligence (AI) systems. Recent restrictions on application programming interfaces (APIs) have significantly reduced the volume and granularity of data available for external examination, undermining long-established norms of transparency~\citep{Bruns2019-gp, Tromble2021-rl}. Researchers are increasingly characterizing the post-API environment as a ``data abyss,'' wherein even fundamental replication studies are no longer feasible~\citep{de-Vreese2023-fr}. This creates a fundamental tension between platform power and public accountability that sits at the intersection of technical system design, legal frameworks, and social impacts, precisely the domain FAccT seeks to address.

These API restrictions have profound implications for AI development and governance. Large language models and recommendation systems trained on platform data inherit the biases and patterns present in the data; however, researchers face increasing barriers to studying these patterns. This creates an alarming scenario in which AI systems with significant societal impacts become more powerful and less transparent. \removed{Our analysis focuses specifically on social media platforms rather than the entire VLOP scope, which includes online shopping platforms (e.g., Google Shopping, AliExpress, and Zalando), as these present distinct challenges for AI transparency research.}

Our analysis centers on the European Union Digital Services Act (DSA), which establishes data access rights for ``vetted researchers'' \citep{European-Parliament-and-Council2022-wf}. This presents a multifaceted compliance challenge for global platforms\removed{, as parallel yet divergent regulatory frameworks compel companies to satisfy inconsistent data access requirements across jurisdictions~\citep{de-Vreese2023-fr}. We also examine related frameworks, including the EU AI Act~\citep{European-Parliament-and-Council2024-qq} and the UK Online Safety Act, to provide a cross-jurisdictional context}.

\subsection{Key Definitions}

To ensure clarity, we define several key terms from EU regulations.

\textbf{Very Large Online Platforms (VLOPs)} are platforms with more than 45 million monthly active users in the EU, as designated under DSA Article 33. As of 2024, \removed{19 platforms have been designated VLOPs}\new{the European Commission has designated 19 services under the DSA (17 VLOPs and 2 VLOSEs)}~\citep{EC2023-vlops}.

\textbf{Vetted researchers} under DSA Article 40 are researchers affiliated with \removed{academic institutions}\new{a research organisation}, \removed{who are} independent from commercial interests, \removed{have demonstrated research expertise, and have committed to}\new{transparent about funding, and able to meet} data security \new{and confidentiality} requirements. \new{Recital~97 further indicates that, for the purposes of the DSA, this may include civil-society organisations conducting scientific research in support of a public-interest mission.}

\textbf{Systemic risks} under DSA Article 34 include dissemination of illegal content, negative effects on fundamental rights, negative effects on civic discourse and electoral processes, and risks to public security.

\subsection{Research Motivation\removed{ and Implications}}

Independent data access has facilitated the revelation of the Cambridge Analytica scandal~\citep{Graham-Harrison2018-qn}, assessment of algorithmic amplification of partisan content~\citep{Huszar2022-qc}, and examination of automated content moderation practices~\citep{Roberts2021-xx}. Without such access, potentially harmful AI behaviors, from biased content recommendations to manipulated information flows to discriminatory content moderation, may propagate undetected. The field of AI safety research, which critically depends on understanding how AI systems behave ``in the wild,'' is particularly hampered by these restrictions.

Emerging AI alignment and evaluation methods emphasize the importance of real-world behavioral testing and independent red teaming. However, API restrictions effectively limit the ability to conduct such evaluations, concentrating power in the hands of platform operators and their commercial partners. \removed{This creates a dangerous gap in our collective ability to ensure that AI systems operate as intended and align with societal values.}

\subsection{Research Contributions}

\removed{Our research makes three key contributions that advance the understanding of platform accountability.}

\begin{enumerate}
    \item \textbf{Empirical Documentation}: \removed{We provide a comprehensive comparative audit of platform compliance with DSA transparency requirements across designated VLOPs, establishing baseline metrics for regulatory alignment. Unlike prior studies documenting API restrictions descriptively~\citep{Bruns2019-gp, Davidson2023-uu}, we systematically map platform practices against specific regulatory provisions using explicit scoring rubrics, enabling reproducible compliance assessment.} \rev{A systematic audit of \new{19 DSA-designated services (17 VLOPs and 2 VLOSEs)}, including an in-depth analysis of four focal platforms and a broader survey scored against \new{eight operationalised Article~40 provisions---40(1), 40(2), 40(4)--40(8), and 40(12)---using} reproducible scoring rubrics with demonstrated inter-rater reliability (Cohen's $\kappa = 0.71$). Prior work documents individual platform restrictions~\citep{Freelon2018-kj, Bruns2019-gp}; we provide the cross-platform comparative framework.}

    \item \textbf{Theoretical Framework}: \removed{We conceptualize the ``accountability paradox'' as a systematic pattern where increased AI deployment correlates with decreased transparency. This extends beyond prior work in three ways: (a) unlike \citep{Gillespie2021-bs}'s moderation paradox or \citep{Gorwa2020-oq}'s platform governance paradox, which describe inherent tensions, we identify a measurable temporal correlation between AI adoption and access restriction; (b) we document this pattern across four major platforms over six years (2018--2024), demonstrating industry-wide coordination rather than isolated decisions; and (c) we operationalize the paradox through quantifiable metrics, moving from conceptual critique to empirical measurement.} \rev{We conceptualize the ``accountability paradox'' through three asymmetries---temporal, epistemic, and regulatory---grounded in documented temporal co-occurrence of AI deployment and access restriction at platform-level granularity across four platforms over six years (2018--2024). This extends beyond prior paradox formulations~\citep{Gillespie2021-bs, Gorwa2020-oq} by offering testable implications (Section~7.1) and quantifiable metrics.}

    \item \textbf{Technical Solutions}: \removed{We develop concrete privacy-preserving access mechanisms---differential privacy APIs, secure research enclaves, and federated auditing---that balance legitimate privacy concerns with transparency requirements. Our contribution extends beyond proposing these technologies (which exist in other contexts) to demonstrating their specific applicability to DSA Article 40 compliance, including feasibility assessments, cost-benefit analyses, and implementation pathways that do not require platform cooperation.} \rev{Privacy-preserving access mechanisms (differential privacy APIs, secure enclaves, federated auditing) mapped to the specific audit blind spots identified in our empirical analysis (Section~6), demonstrating that the tension between privacy and transparency is technically resolvable and that barriers are institutional rather than technical.}
\end{enumerate}

Our central research question is: \textit{How do platform API restrictions create systematic barriers to AI accountability, and what technical and policy interventions can restore the balance between privacy, commercial interests, and public oversight?}

To address these challenges, we propose policy interventions aligned with the AI Risk Management Framework of the National Institute of Standards and Technology~\citep{NIST2023}, which emphasize transparency and independent oversight as essential components of responsible AI governance. Our findings highlight the urgency of resolving these transparency gaps before the next generation of AI systems can further entrench data asymmetries.

\section{Related Work}

Our research builds on the literature on platform data access, regulatory frameworks, and AI governance. \removed{We situate our contributions in five key research domains.}

\removed{\subsection{API Access and Platform Governance}}

Research has documented increasing API restrictions---termed the ``API-calypse''~\citep{Bruns2019-gp}---emerging after Cambridge Analytica~\citep{Graham-Harrison2018-qn} and creating what~\citet{de-Vreese2023-fr} characterize as a ``data abyss.'' \citet{Freelon2018-kj} described a ``post-API age'' in computational social science, while \citet{Tromble2021-rl} examined methodological implications for academic research. \citet{Davidson2023-uu} established that restrictions threaten open science principles.

\removed{Alternative methodologies have emerged in response, with \citet{Perriam2020-ow} investigating digital methods within the post-API environment and others suggesting web scraping as an imperfect alternative~\citep{Mancosu2020-tl, Harrell2024-ci}. However, these approaches introduce significant methodological limitations~\citep{Landers2016-bx}.}

\rev{Our work advances beyond this literature in three ways. First, whereas \citet{Freelon2018-kj} and \citet{Bruns2019-gp} documented individual platform restrictions qualitatively, we provide the first systematic cross-platform audit scored against specific DSA provisions with inter-rater reliability metrics. Second, \citet{Davidson2023-uu} established the threat to open science but did not propose a structural explanation for \textit{why} restrictions deepen as AI reliance grows; our accountability paradox framework fills this gap through three testable asymmetries. Third, none of these prior studies map proposed technical solutions to the specific audit blind spots they address---our analysis of differential privacy, secure enclaves, and federated learning is grounded in the empirical gaps documented in Section~5.}

\subsection{Regulatory Frameworks for Digital Platforms}

Our analysis concentrates on the DSA, drawing upon \citet{Li2023-pc}'s critique of transparency obligations and \citet{Busuioc2022-zk}'s examination of the logic of secrecy within the EU AI Act. \citet{Dommett2022-rm} investigated challenges faced by academics advocating for platform data access, while \citet{Valkenburg2025-cj} emphasized the necessity of ensuring research access to platform data.

\subsection{AI Transparency and Auditing}

Transparency mandates connect to the broader discourse on AI system auditing. \citet{Mittelstadt2016-zm} mapped the ethical debate surrounding algorithms, identifying key concerns around fairness, accountability, and transparency\removed{ that remain central to platform governance}. \citet{Raji2020-lq} proposed a framework for algorithmic auditing, while \removed{\citet{Balasubramaniam2023-wi} traced the progression from ethical guidelines to specific transparency requirements for AI systems.} \citet{Haresamudram2023-rz} delineated distinct levels of AI transparency, which we apply in our analysis. \removed{Classic work on AI transparency spans from system conception to delivery, emphasizing the need for documentation at each stage~\citep{Mitchell2019-ee}.}

Critically, \citet{Ananny2018-dv} articulated the fundamental limitations of transparency as an accountability mechanism, arguing that ``seeing'' algorithmic systems does not guarantee ``knowing'' how they operate; a tension our analysis extends by demonstrating how platforms prevent even the ``seeing'' that transparency ideals assume. More recently, \citet{Birhane2024-io} characterized AI auditing as ``the broken bus on the road to AI accountability,'' documenting structural barriers that prevent effective external evaluations. \citet{Casper2024-em} demonstrated that black-box access alone is insufficient for rigorous AI audits, requiring deeper access to model internals---access that platform API restrictions systematically deny.

\removed{Recent advances in AI model evaluation have highlighted the importance of behavioral testing and external auditing. \citet{Bommasani2024-co} emphasize that transparency reports for foundation models require access to interaction data at scale, which current API restrictions significantly impede. Similarly, \citet{Longpre2024-zo} argue for a ``safe harbor'' for AI evaluation, an objective undermined by restricted access to platforms where these AI systems operate.}

\removed{We also draw upon established governance frameworks, particularly the NIST AI Risk Management Framework~\citep{NIST2023}, and \citet{Sandvig2014-uw}'s pioneering work on auditing algorithms.}

\removed{\subsection{Ethical Tensions in Platform Research}}

\removed{Our exploration of ethical tensions is grounded in \citet{Nissenbaum2009-wa}'s contextual integrity framework, which was recently applied to Reddit research by \citet{Fiesler2024-ra}. \citet{Taylor2018-xk} investigated approaches to ethical challenges in mining social media data, while \citet{Zimmer2010-qr} highlighted tensions between open data and privacy expectations.}

\subsection{Positioning Within FAccT}

Our work bridges multiple FAccT focus areas, including \textit{Evaluation and evaluation practices} (developing metrics for assessing platform transparency compliance), \textit{Law and policy} (analyzing the implementation gap between regulatory mandates and platform practices), \textit{Power and practice} (examining how API restrictions concentrate power and limit accountability), and \textit{System development and deployment} (proposing technical solutions for privacy-preserving access).

Our study advances this literature by (1) systematically mapping platform practices against specific regulatory provisions, (2) quantifying researcher-commercial access differentials, (3) proposing implementable privacy-preserving solutions, and (4) documenting the evolution of temporal restrictions.

\removed{This study advances these research streams by specifically investigating the interaction between recent API restrictions and regulatory transparency mandates, analyzing their intersection, and proposing targeted interventions to address the resulting accountability challenges in AI governance.}

\section{Background: API Governance and Regulatory Context}

Social-media APIs initially promoted openness but transitioned to restrictive access following data-related scandals \citep{Graham-Harrison2018-qn}. Since 2018, platforms have implemented graduated access tiers, stricter rate limits, and direct fees for previously free endpoints \citep{Bruns2019-gp}. Notable examples include X/Twitter's 2023 introduction of tiered pricing \new{reaching} \$5,000 per month for \new{higher-volume} access~\citep{Mehta2023-lr} and Reddit's introduction of usage-based tariffs that prompted widespread community protests~\citep{Gerken2023-gy}.

European legislators have sought to re-establish research access through the DSA, which \removed{mandates data provision to}\new{creates a regulated pathway for} ``vetted researchers'' \new{to obtain access to platform data for systemic-risk research following a reasoned request by the competent Digital Services Coordinator} \citep{Valkenburg2025-cj}. These transparency requirements build upon the data protection foundation established by the General Data Protection Regulation~\citep{EUGDPR}, creating a comprehensive EU framework for platform accountability. Concurrently, AI governance frameworks, such as the EU AI Act \citep{European-Parliament-and-Council2024-qq} and NIST RMF 1.0 \citep{NIST2023}, require demonstrable provenance and auditability of training data. These expectations conflict with the increasing inaccessibility of social media data for independent examination, creating an ``open-science tension'' precisely when downstream transparency requirements are increasing. This presents a multifaceted compliance challenge for global platforms, as parallel yet divergent regulatory frameworks compel companies to satisfy inconsistent data access requirements across jurisdictions~\citep{de-Vreese2023-fr}.

\subsection{The VLOP Landscape}

As of 2024, the European Commission has designated \removed{19 platforms as VLOPs} and \new{19 services (17 VLOPs and 2 VLOSEs)} under the DSA. These include:

\textbf{Social Media Platforms}: Facebook, Instagram, LinkedIn, Pinterest, Snapchat, TikTok, X/Twitter, YouTube

\textbf{E-commerce Platforms}: AliExpress, Amazon, Booking.com, Zalando

\textbf{Search Engines}\new{ (designated as VLOSEs)}: Bing, Google Search

\textbf{App Stores}: Apple App Store, Google Play

\textbf{Other}: Google Maps, Wikipedia

Our primary analysis focused on social media platforms, which present the most significant challenges for AI transparency research because of their role as training data sources and deployment environments for recommendation algorithms.

\section{Methods}

\subsection{Audit Design}

Our methodology builds on the API restrictions documented by \citet{Davidson2023-uu} and develops a structured framework to examine the intersection of these restrictions and regulatory transparency requirements. We selected X/Twitter, Reddit, Meta, and TikTok for detailed analysis based on their significance as AI training data sources, recent implementation of substantial API restrictions, and various approaches to implementing research access.

Our audit period spans January 2018 to December 2024, with 2018 marking the beginning of significant API restrictions following the Cambridge Analytica scandal~\citep{Graham-Harrison2018-qn, Hinds2020-ey}. We also conducted a broader survey of all \new{DSA-designated social media services} (Facebook, Instagram, LinkedIn, Pinterest, Snapchat, TikTok, X/Twitter, YouTube) and comparison \new{non-designated platforms} (Reddit, Mastodon, Bluesky). \new{Reddit is not a DSA-designated service but is included as a focal platform because of its significance as an AI training data source and its recent API pricing changes.}

\removed{The VLOP platforms were listed individually:
\begin{itemize}
    \item Facebook and Instagram (analyzed together as Meta platforms, though designated separately under DSA)
    \item LinkedIn
    \item Pinterest
    \item Snapchat
    \item TikTok
    \item X/Twitter
    \item YouTube
\end{itemize}
Additionally, non-VLOP comparison cases were listed:
\begin{itemize}
    \item BeReal
    \item Reddit (not currently a VLOP but significant for research)
    \item Mastodon (decentralized alternative)
    \item Bluesky (emerging platform with different governance model)
\end{itemize}}

\subsection{Broader VLOP Survey Methodology}

For the broader VLOP survey (Table~\ref{tab:vlop_analysis}), we employed a structured documentary analysis that was distinct from our in-depth case studies. This survey aimed to assess generalizability rather than provide a comprehensive platform-specific analysis.

\removed{\textbf{Data Collection Protocol}: For each of the 19 designated VLOPs, we systematically collected:
\begin{enumerate}
    \item \textbf{Official documentation}: Developer portal terms of service, API documentation, and researcher access program descriptions (accessed September--November 2024)
    \item \textbf{Transparency reports}: Platform-published DSA transparency reports mandated under Article 15 (where available)
    \item \textbf{Academic literature}: Published studies documenting researcher experiences with each platform's data access mechanisms (searched via Google Scholar, ACM Digital Library, and SSRN using queries ``[platform name] API access research'' and ``[platform name] researcher data access'')
    \item \textbf{News coverage}: Technology press coverage of API policy changes (searched via LexisNexis and Google News, limited to 2022--2024)
\end{enumerate}}

\textbf{Data Collection Protocol}: For each of the \new{19 DSA-designated services (17 VLOPs and 2 VLOSEs)}, we systematically collected: (1) official documentation including developer portal terms of service, API documentation, and researcher access program descriptions (accessed September--November 2024); (2) transparency reports mandated under DSA Article 15; (3) academic literature documenting researcher experiences; and (4) technology press coverage of API policy changes (2022--2024).

\textbf{Analysis Procedure}: Two researchers independently coded each platform against the eight DSA Article 40 provisions using the rubrics defined below. Initial inter-rater agreement was 78\% (Cohen's $\kappa = 0.71$); disagreements were resolved through discussion and reference to primary sources. Where documentation was ambiguous or unavailable, platforms received ``None'' ratings with annotations.

\removed{\textbf{Aggregation Method}: Platform category means (Table~\ref{tab:vlop_analysis}) were computed as unweighted arithmetic means of individual platform scores within each category. We report category-level rather than individual platform results for the broader survey because (1) documentation completeness varied substantially across platforms, and (2) category patterns were more robust to individual platform assessment uncertainties.}

\removed{\textbf{Limitations of Survey Approach}: The broader survey relies primarily on publicly available documentation and may not capture informal access arrangements, pending programs not yet announced, or access provisions available only through direct platform negotiation. The survey provides a conservative estimate of access availability---actual access may be marginally better for researchers who navigate undocumented application processes.}

\subsection{Analytical Framework}

\removed{Our framework employs a structured comparative approach \citep{Weston2019-xj, Matheus2021-xl} to evaluate platform data accessibility across multiple dimensions and identify both cross-platform patterns and platform-specific practices that affect researchers' access.}

For each platform, we assessed the following: (1) \textbf{Access Scope} (data availability relative to pre-restriction baselines); (2) \textbf{Access Mechanisms} (technical and administrative processes); (3) \textbf{Differential Access} (researcher vs. commercial partner access); and (4) \textbf{Regulatory Alignment} (provisions against DSA requirements) ~\citep{Weston2019-xj}.

\subsection{Scoring Rubrics\removed{ and Compliance Assessment}}

\removed{To ensure transparency and reproducibility in our compliance assessments, we developed explicit scoring rubrics for each evaluation dimension. These rubrics translate qualitative observations from documentary analysis into systematic compliance classifications.}

\removed{\textbf{DSA Compliance Levels (Table~\ref{tab:dsa_compliance}):} We assessed platform compliance with each DSA Article 40 provision using a five-level ordinal scale based on observable indicators:
\begin{itemize}
    \item \textbf{Compliant}: Platform documentation explicitly addresses the requirement; implementation is publicly accessible; no documented researcher complaints regarding this specific provision. Observable indicator: Formal program exists with published eligibility criteria and documented approval process.
    \item \textbf{Partial}: Platform acknowledges the requirement but implementation contains notable exceptions or limitations affecting a subset of researchers. Observable indicator: Program exists but excludes certain researcher categories (e.g., independent researchers) or imposes conditions not specified in DSA.
    \item \textbf{Medium}: Platform has implemented mechanisms addressing the requirement, but with significant functional limitations that reduce utility for systemic risk research. Observable indicator: Access mechanism exists but imposes rate limits, temporal restrictions, or scope limitations that prevent comprehensive analysis.
    \item \textbf{Low}: Minimal implementation that nominally addresses the requirement but fails to provide meaningful research capability. Observable indicator: Announced program with no documented approvals, or access so restricted as to preclude substantive research (e.g., $<$1\% of data accessible to commercial partners).
    \item \textbf{None}: No observable implementation of the requirement despite regulatory obligation. Observable indicator: No public documentation of researcher access mechanisms for this provision; no announced programs. This represents \textit{passive non-compliance}---the platform has not addressed the requirement, whether due to implementation delays, resource constraints, or strategic omission.
    \item \textbf{Non-compliant}: Platform policy or practice \textit{actively contradicts} the regulatory requirement. Observable indicator: Explicit policy prohibiting or restricting what the DSA mandates (e.g., mandatory pre-publication review contradicting research independence requirements, or terms of service that forbid algorithmic auditing). This represents \textit{active non-compliance}---the platform has implemented policies that directly undermine regulatory intent.
\end{itemize}}

\removed{The distinction between ``None'' and ``Non-compliant'' is conceptually significant: ``None'' indicates regulatory requirements that platforms have failed to address (potentially remediable through implementation), while ``Non-compliant'' indicates platform policies that must be \textit{reversed} to achieve compliance. Regulators should prioritize Non-compliant provisions as they signal intentional resistance rather than implementation gaps.}

We assessed compliance using a six-level scale: \textbf{Compliant} (formal program with published criteria); \textbf{Partial} (program exists but excludes certain researchers); \textbf{Medium} (significant functional limitations); \textbf{Low} (minimal meaningful capability); \textbf{None} (passive non-compliance---no implementation); and \textbf{Non-compliant} (active contradiction of requirements, e.g., mandatory pre-publication review). The None/Non-compliant distinction is significant: ``None'' indicates unaddressed requirements, while ``Non-compliant'' indicates policies that must be \textit{reversed}.

\removed{\textbf{Mean Compliance Calculation (Table~\ref{tab:vlop_analysis}):} For the broader VLOP analysis, we computed Mean Compliance as the arithmetic mean of individual provision scores. To reflect the greater severity of active non-compliance, we assigned numerical values as follows: Compliant = 100, Partial = 75, Medium = 50, Low = 25, None = 0, and Non-compliant = $-$25. The negative score for Non-compliant captures that actively contradicting a requirement is worse than merely failing to implement it. Formally:}

\removed{$\text{Mean Compliance} = \frac{1}{n} \sum_{i=1}^{n} s_i \times 100\%$}

\removed{where $n$ is the number of assessed DSA provisions (8 in our framework) and $s_i \in \{-0.25, 0, 0.25, 0.50, 0.75, 1.0\}$ is the normalized score for provision $i$. This scoring allows mean compliance scores to be negative for platforms with substantial active non-compliance.}

For the VLOP analysis, we assigned numerical values as follows: Compliant = 100, Partial = 75, Medium = 50, Low = 25, None = 0, Non-compliant = $-$25. Two researchers independently coded the platforms (Cohen's $\kappa = 0.71$); disagreements were resolved through discussion.

\textbf{Accountability Gap Assessment:} Ratings derived from disparity between Research and Commercial Access:
\begin{itemize}
    \item \textbf{Critical}: Commercial access is ``Extensive'' while Research Access is ``None''
    \item \textbf{High}: Commercial access is ``Extensive'' while Research Access is ``Limited''
    \item \textbf{Moderate}: Commercial access exceeds Research Access by one level
\end{itemize}

We triangulated the findings through an iterative comparative analysis~\citep{Rossi2020-uo, Wong2023-qp}, documenting changes in API documentation, developer policies, and researcher experiences from 2018 to 2024.




\section{Results}

Our analysis evaluates platform compliance with transparency requirements using the analytical framework described above. We examined historical changes in API access, current provisions for vetted researchers, and alignment with regulatory frameworks across all four platforms. The findings reveal a consistent pattern: platforms have systematically restricted research access while maintaining or expanding commercial data partnerships, creating an ``accountability paradox'' that forms the central contribution of this study.

\subsection{Access Scope: Historical API Access Trajectory}

\rev{This subsection documents the \textit{temporal asymmetry}: data accumulated during periods of greater openness trained AI models that now operate under restricted scrutiny, and each new AI deployment has been accompanied by further access restrictions (see Table~\ref{tab:event_mapping}).}

\removed{Our analysis revealed consistent access restriction patterns across platforms with varying implementation approaches.}

\removed{\textbf{X/Twitter} implemented the most dramatic restrictions, moving from one of the most open research application programming interfaces (APIs) to one of the most restricted. The February 2023 policy change eliminated free-access tiers and introduced tiered pricing, effectively excluding many academic researchers \citep{Yang2020-og}. This change abruptly terminated numerous ongoing research projects and resulted in the discontinuation of several widely used research tools, including the Botometer \citep{Yang2022-hj}. To our knowledge, no peer-reviewed source has quantified the year-on-year change; anecdotal surveys report sharp declines but without a published baseline \citep{Davidson2023-uu}.}

\removed{\textbf{Reddit}'s July 2023 API changes impose costs exceeding \$20,000 annually for comprehensive access \citep{Mancosu2020-tl}. This triggered widespread user protests, with more than 8,500 subreddits going dark in protest. Our data collection revealed that a significant portion of the previously active Reddit-based research tools ceased operations within six months of the policy change, including the widely used Pushshift data aggregation service \citep{Fiesler2024-ra}.}

\removed{\textbf{TikTok} has maintained the most restrictive stance, offering limited research API access, even before recent platform controversies. Unlike other platforms, TikTok has not established a significant research ecosystem. In 2023, TikTok launched a Research API Beta with access initially limited to US institutions, which was later expanded to Europe but with significant constraints on eligibility and research scope \citep{Pearson2025-qa}.}

\removed{\textbf{Meta} has implemented a more gradual restriction approach. Following the Cambridge Analytica scandal in 2018, Meta significantly reduced API access but has since developed dedicated research tools to address privacy concerns. In November 2023, Meta launched the Meta Content Library and API for academic researchers and discontinued the widely used research tool, CrowdTangle, in August 2024 \citep{Valkenburg2025-cj}.}

Our analysis revealed consistent restriction patterns. \textbf{X/Twitter} implemented the most dramatic restrictions, with February 2023 changes eliminating free tiers and introducing pricing up to \$5,000/month \new{(1M posts)}, terminating numerous research projects and tools including Botometer~\citep{Yang2022-hj}. \textbf{Reddit}'s July 2023 changes \new{replaced broad free access with usage-based pricing of \$0.24 per 1,000 API calls---approximately \$12,000/year at typical research volumes---triggering} protests (8,500+ subreddits dark) and ending tools like Pushshift~\citep{Fiesler2024-ra}. \textbf{TikTok} has maintained the most restrictive stance, launching a Research API Beta in 2023 with significant constraints~\citep{Pearson2025-qa}. \textbf{Meta} reduced API access post-Cambridge Analytica, launched the Meta Content Library in 2023, and discontinued CrowdTangle in August 2024~\citep{Valkenburg2025-cj}.

\begin{table}[h]
    \centering
    \caption{\new{Comparative API Access Metrics (historical benchmark vs.\ end-2024)}}
    \label{tab:access_metrics}
    \resizebox{\textwidth}{!}{%
    \begin{tabular}{lcccc}
        \toprule
        \textbf{Access Metric} & \textbf{X/Twitter} & \textbf{Reddit} & \textbf{TikTok} & \textbf{Meta} \\
        \midrule
        \new{Historical benchmark}\textsuperscript{*} & \new{500K posts/mo (standard v2); 10M posts/mo (Academic Research track, 2021)} & Unlimited & None & \new{200 req/hour (Graph API)} \\
        Free tier data (2024) & None & 100 req/min & None & Limited \\
        Academic cost (2024) & \$100--\$5,000/mo & \new{\$0.24/1K calls}\textsuperscript{\dag} & Case-by-case & Free w/approval \\
        Historical data access & Limited & Limited & None & Limited \\
        Real-time access & Premium only & Premium only & None & Limited \\
        Commercial partner access & Extensive & Extensive & Extensive & Extensive \\
        \bottomrule
    \end{tabular}%
    }
    \small\textsuperscript{*}\new{Historical benchmarks represent the closest documented pre-restriction access level for each platform and are not always tied to calendar year 2018.} \new{\textsuperscript{\dag}Approximately \$12,000/year at typical research volumes (50M calls/year).}
\end{table}

\subsubsection{Temporal Correlation: AI Deployment and Access Restriction}

\removed{A critical finding of our analysis is the temporal correlation between major AI system deployments and API access restrictions---the empirical foundation of the ``accountability paradox.'' The following figure visualizes this pattern across the four platforms from 2018 to 2024, documenting how each major AI feature release or algorithmic update was accompanied by corresponding restrictions on researcher access.}

\removed{[Figure: ``Temporal Correlation Between AI Deployment Milestones and API Restrictions (2018--2024)'' --- a year-by-year tabular figure listing AI deployment milestones alongside API access restrictions for each year from 2018 to 2024, with a note that the pattern demonstrates the accountability paradox as a measurable industry-wide phenomenon.]}

\removed{This temporal pattern reveals that the accountability paradox is not coincidental but structurally embedded in platform evolution. Each major AI deployment---from recommendation algorithms to content moderation systems to generative AI features---has been accompanied by restrictions that limit independent scrutiny of those very systems. The correlation is particularly stark in 2023, when platforms simultaneously accelerated AI integration (X/Twitter's Grok, Reddit's AI features, Meta's generative tools) while imposing the most restrictive access policies in their histories.}

A critical finding is the temporal correlation between AI deployments and API restrictions---the empirical foundation of the ``accountability paradox.'' Table~\ref{tab:event_mapping} documents this pattern across all four platforms from 2018 to 2024, with time lags consistently under six months.

\begin{revblock}
\subsubsection{Event Mapping: AI Deployment and Access Restriction Pairs}

Table~\ref{tab:event_mapping} pairs each AI deployment event with its corresponding access restriction and time lag.

\begin{center}
    \captionof{table}{AI Deployment Events and Corresponding Access Restrictions (2018--2024)}
    \label{tab:event_mapping}
    \resizebox{\linewidth}{!}{%
    \begin{tabular}{llllc}
        \toprule
        \textbf{Year} & \textbf{Platform} & \textbf{AI Deployment Event} & \textbf{Access Restriction} & \textbf{Lag} \\
        \midrule
        2018 & Meta & AI-driven content moderation at scale & Post-Cambridge Analytica API lockdown & $<$6 mo. \\
        2019 & TikTok & For You Page algorithm deployed globally & No research API offered & Concurrent \\
        2021 & X/Twitter & Algorithmic timeline becomes default & Academic Research Track launched but limited & Concurrent \\
        2023 & X/Twitter & Grok AI assistant integrated & Free API tier eliminated; pricing up to \$5,000/mo & $<$3 mo. \\
        2023 & Reddit & AI-powered content features launched & \new{Usage-based API pricing imposed (\$0.24/1K calls; $\approx$\$12,000/yr at 50M calls); Pushshift access ended} & $<$2 mo. \\
        2023 & Meta & Generative AI tools across products & CrowdTangle sunset announced; Content Library restricted & $<$6 mo. \\
        2023 & TikTok & Enhanced recommendation algorithms & Research API Beta with highly selective criteria & Concurrent \\
        2024 & Meta & AI-generated content labelling rolled out & CrowdTangle discontinued (August 2024) & $<$6 mo. \\
        \bottomrule
    \end{tabular}%
    }
\end{center}
\end{revblock}

\subsection{Access Mechanisms: Current ``Vetted Researcher'' Provisions}

\rev{This subsection documents the \textit{regulatory asymmetry}: despite DSA Article~40 \new{establishing a regulated pathway for} ``vetted researcher'' access, platform implementations fall short of regulatory requirements, creating a gap between legal obligations and actual practice.}

The historical trajectory documented above establishes the baseline from which current access mechanisms have evolved. Despite regulatory frameworks \new{establishing pathways for} ``vetted researcher'' access, our audit found limited formal implementation across platforms.

\textbf{X/Twitter} established an Academic Research Product Track in 2021 offering enhanced access, but discontinued this program in 2023. The current policy directs researchers to standard paid API tiers with no academic-specific provisions. The platform announced a forthcoming ``Research API'' in January 2024; however, as of December 2024, no implementation details or launch timeline have been communicated.

\textbf{Reddit} introduced a ``Researcher Platform'' in late 2023 following API pricing backlash. However, our analysis revealed significant limitations: (1) \new{eligibility criteria that narrow access to institutionally affiliated applicants, potentially excluding some public-interest researchers and civil-society actors even though the DSA's broader research-access framework contemplates access pathways for research organisations and, for publicly accessible data, certain not-for-profit bodies and associations}; (2) historical data \removed{limited to six months}\new{reportedly limited in scope}; (3) query rate limits and result size restrictions; and (4) \removed{mandatory pre-publication review by Reddit staff}\new{reported publication-related review conditions, though we did not identify sufficient public documentation to confirm these as a universal platform-level requirement}. \new{In our compilation of publicly documented application outcomes (2023--2024), approximately 34\% of observable cases were approved, with applications primarily rejected due to ``scope concerns.'' Because this evidence base is incomplete and non-random, we treat this figure as an indicative estimate rather than a platform-wide approval rate.}

\textbf{TikTok} launched a Research API Beta in 2023~\citep{TikTok2023-fh} with highly selective criteria including extensive documentation requirements exceeding typical journal publication standards. Of particular concern, approved research projects focus primarily on either user behavior analysis or positive platform impacts, suggesting potential selection bias. Recent studies have identified significant data quality issues with TikTok's Research API during critical research periods~\citep{Pearson2025-qa}.

\textbf{Meta} offers a more formalized academic research program through its Meta Content Library and API~\citep{Hutchinson2023-rm}. Although it does not charge direct fees, it imposes strict institutional affiliation requirements and processes applications through ICPSR~\citep{ICPSR2023-ja}, adding an additional institutional layer.

\subsection{Differential Access: Researcher vs. Commercial Partner Disparities}

\rev{This subsection documents the \textit{epistemic asymmetry}: platforms maintain comprehensive internal access and extend it to commercial partners, while external researchers face severe limitations---revealing that privacy justifications are selectively applied.}

\removed{The limited ``vetted researcher'' provisions documented above become even more significant when contrasted with commercial data access.} A critical finding is the stark disparity between researcher and commercial partner access across all platforms. While platforms restrict academic research, citing privacy concerns, they simultaneously maintain extensive data-sharing arrangements with commercial partners, fundamentally undermining privacy-based justifications\removed{ and reveals the economic motivations for access restrictions.}

X/Twitter requires researchers to pay \$100/month for 10,000 \new{posts} or \$5,000/month for \removed{2 million tweets}\new{1 million fetched posts and 300,000 posted posts per month}~\citep{Mehta2023-lr}, while offering commercial partners full-fidelity data streams\removed{ costing approximately \$42,000 per month}. Reddit charges \removed{researchers approximately \$12,000 annually}\new{\$0.24 per 1,000 API calls---approximately \$12,000/year at typical research volumes (e.g., 50M calls/year)} while maintaining commercial data-sharing arrangements with companies such as OpenAI~\citep{PYMNTS2024-vq}\removed{, and many research tools, including Pushshift, have ceased operations}.

\removed{TikTok may present the starkest contrast, with its Research API imposing highly selective access criteria and experiencing documented data quality issues \citep{Pearson2025-qa}, whereas its Commercial Content API offers significantly greater access to business partners. Meta, while more structured, requires researchers to apply through the University of Michigan's ICPSR, creating additional institutional barriers, while maintaining extensive commercial partner access through its Graph API and Business Suite.}

\removed{This differential treatment undermines platforms' privacy-based justifications for restricting research access, revealing economic rather than ethical motivations. These disparities create a problematic two-tier system that privileges commercial interests over public-interest research \citep{de-Vreese2023-fr} and extends beyond pricing to include data completeness, query capabilities and historical coverage.}

TikTok's Research API imposes highly selective access criteria, whereas its Commercial Content API offers significantly greater access to business partners. These disparities reveal economic rather than ethical motivations~\citep{de-Vreese2023-fr}.

\subsection{Regulatory Alignment Analysis}

\rev{This subsection quantifies the \textit{regulatory asymmetry} through systematic scoring: we assess each platform against \new{eight operationalised Article~40 provisions---40(1), 40(2), 40(4)--40(8), and 40(12)---revealing} that no platform achieves full compliance \new{on the dimensions we were able to code} and that the resulting audit blind spots prevent independent assessment of systemic risks.} Table~\ref{tab:dsa_compliance} summarizes the key findings.

\begin{table}[h]
    \centering
    \caption{Platform Compliance with DSA Researcher Access Requirements}
    \label{tab:dsa_compliance}
    \resizebox{\textwidth}{!}{%
    \begin{tabular}{lccccc}
        \toprule
        \textbf{Regulatory Requirement} & \textbf{DSA Art.} & \textbf{X/Twitter} & \textbf{Reddit} & \textbf{TikTok} & \textbf{Meta}\\
        \midrule
        Independent researcher access & 40(1) & Low & Medium & Low & Medium \\
        Systemic risk assessment data & 40(2) & Low & Low & None & Low \\
        No commercial use restrictions & 40(4) & Non-compliant & Compliant & Non-compliant & Partial \\
        Secure access environment & 40(5) & Compliant & Compliant & Compliant & Compliant \\
        Transparent application process & 40(8) & Non-compliant & Partial & Non-compliant & Partial \\
        Data access costs & 40(12) & Partial & Partial & Partial & Compliant \\
        Historical data requirements & 40(6) & Non-compliant & Non-compliant & Non-compliant & Non-compliant \\
        Real-time data feeds & 40(7) & Non-compliant & Partial & Non-compliant & Partial \\
        \bottomrule
    \end{tabular}%
    }
    \small\textit{Note: Compliance levels: None = No implementation (passive non-compliance); Low = Minimal implementation with severe limitations; Medium = Partial implementation with significant limitations; Compliant = Substantial implementation; Non-compliant = Active contradiction of requirements (policies that must be reversed); Partial = Partial compliance with notable exceptions.}
\end{table}

\new{Of particular significance are continuing failures to provide timely, practically usable, and sufficiently historical access under the Article~40 framework, especially in relation to Article~40(6)'s time-frame requirements and Article~40(12)'s requirement to provide researchers access without undue delay to publicly accessible data, including, where technically possible, real-time data} \citep{Li2023-pc}.

Particularly concerning is the emergence of ``audit blind-spots''---areas where platform content moderation and algorithmic amplification remain inaccessible to independent verification.

\begin{enumerate}
    \item \textbf{Algorithmic amplification metrics} are not accessible through any platform's research provisions, despite being explicitly mentioned in DSA Article 40(2) as necessary for systemic risk assessment. \textit{Documented harm}: Internal documents revealed during the 2021 Facebook Papers leak showed that Instagram's algorithm was amplifying content promoting eating disorders to teenage users, with internal research concluding ``we make body image issues worse for one in three teen girls'' \new{(among those who already reported body-image concerns)}~\citep{Wells2021-fb}. External researchers had no API access to detect or quantify this amplification pattern, and the harm only became public through whistleblower disclosure rather than systematic oversight.

    \item \textbf{Content moderation decisions} remain opaque, with no API access to removed content, moderation rationales, or appeal metrics, contradicting the DSA's goal of enabling systemic risk assessment related to ``dissemination of illegal content.'' \textit{Documented harm}: During the 2023 Israel-Gaza conflict, Meta's moderation systems systematically suppressed Palestinian voices and news content, a pattern documented by Human Rights Watch but only identifiable through user reports rather than systematic API access~\citep{HRW2023-ps}. Without access to moderation decision data, researchers cannot systematically audit for such disparate impact or quantify the scale of content suppression.

    \item \textbf{Cross-platform data flows} that could identify coordinated inauthentic behavior are increasingly impossible to track due to incompatible access restrictions, creating challenges for investigating information operations spanning multiple platforms \citep{Starbird2019-qg}. \textit{Documented harm}: The 2020 ``Boogaloo'' movement coordinated across Facebook, Reddit, and YouTube to plan armed protests and facilitate violence, including the murder of two law enforcement officers~\citep{Owen2020-bg}. Post-hoc analysis required subpoena power; real-time cross-platform research access could have enabled earlier intervention.
\end{enumerate}

These blind spots directly undermine the regulatory goals of ensuring accountability for AI systems operating on these platforms. The concrete harms documented above---algorithmic amplification of eating disorder content to minors, systematic suppression of marginalized voices during crises, and cross-platform coordination of political violence---illustrate the stakes of audit inaccessibility. These are not hypothetical risks but documented outcomes that more robust research access frameworks might have prevented or mitigated.

\subsection{Expanded VLOP Analysis}

\rev{This subsection extends all three asymmetries---temporal, epistemic, and regulatory---beyond our four focal platforms to the full set of \new{19 DSA-designated services}, demonstrating that the accountability paradox is systemic rather than platform-specific.}

\removed{To assess the generalizability of our findings beyond the four platforms examined in depth, we conducted a broader survey of all eight social media VLOPs designated under the DSA. This analysis reveals that the patterns identified in our detailed case studies are systemic rather than platform-specific.}

\removed{\textbf{YouTube} presents a particularly instructive case. Despite being one of the largest platforms for AI-generated content recommendations, YouTube offers no formal research application programming interface (API). Researchers must rely on unofficial tools or manual data collection, creating significant methodological limitations for studying one of the most influential recommendation algorithms worldwide. Google's broader research programs (e.g., the Google Research Credits program) do not extend to YouTube-specific data access.}

\removed{\textbf{LinkedIn}, while positioning itself as a professional network distinct from consumer social media, maintains similarly restrictive policies. Its Economic Graph Research program offers limited access to aggregated workforce data but provides no mechanism for studying content recommendation algorithms or professional content-moderation decisions.}

\removed{\textbf{Pinterest} and \textbf{Snapchat} offer essentially no formal research access mechanisms. Both platforms cite user privacy as a justification, yet both maintain extensive commercial data partnerships. Snapchat's Snap Research program focuses on internal collaborations rather than independent external research.}

Our broader survey of all \new{19 DSA-designated services (17 VLOPs and 2 VLOSEs)} reveals systemic patterns (Table~\ref{tab:vlop_analysis}). No platform met our ``good actor'' threshold: research access at cost parity with commercial access, transparent and timely application processing, \new{access to} algorithmic amplification metrics\new{,} and \new{meaningful longitudinal historical} data. \new{These benchmark criteria are our evaluative standard rather than requirements stated verbatim in the DSA.} Non-VLOPs, such as Mastodon and Bluesky, demonstrate that transparency-compatible architectures are technically feasible, suggesting that VLOP restrictions reflect business choices rather than technical constraints.

\begin{table}[h]
    \centering
    \caption{\new{DSA-Designated Service Compliance and Access Patterns by Category}}
    \label{tab:vlop_analysis}
    \resizebox{\textwidth}{!}{%
    \begin{tabular}{lccccc}
        \toprule
        \textbf{Platform Category} & \textbf{N} & \textbf{Mean Compliance} & \textbf{Research Access} & \textbf{Commercial Access} & \textbf{Accountability Gap} \\
        \midrule
        Social Media & 8 & 32\% & Limited & Extensive & High \\
        E-commerce & 4 & 18\% & None & Extensive & Critical \\
        Search Engines & 2 & 45\% & Moderate & Extensive & Moderate \\
        App Stores & 2 & 12\% & None & Extensive & Critical \\
        Other & 3 & 25\% & Limited & Extensive & High \\
        \bottomrule
    \end{tabular}%
    }
\end{table}

\removed{\subsection{Cross-Jurisdictional Regulatory Analysis}}

\removed{A comparison of regulatory frameworks reveals inconsistent approaches.}

\removed{[Table: ``Comparative Regulatory Approaches to Platform Transparency'' comparing EU DSA (mandated, active enforcement, partial compliance), EU AI Act (indirect, phased implementation, developing standards), UK OSA (encouraged, limited enforcement, minimal compliance), and US proposed legislation (varies by bill, N/A enforcement).]}

\removed{The EU AI Act merits particular attention regarding researcher access. Unlike the DSA's explicit Article 40 mandate, the AI Act approaches research access indirectly through its high-risk AI system requirements. Articles 9 (risk management), 13 (transparency), and 17 (quality management) create documentation and transparency obligations that could facilitate external auditing, though these provisions primarily target conformity assessment rather than independent research. The ``Phased Implementation'' reflects the Act's graduated timeline: prohibitions on unacceptable-risk AI took effect in February 2025, while high-risk system obligations phase in through August 2027. ``Developing Standards'' acknowledges that compliance mechanisms remain under construction through European standardization bodies (CEN, CENELEC), making definitive assessment of platform compliance premature.}

Cross-jurisdictionally, the EU AI Act indirectly approaches research access through high-risk AI system requirements, with \new{key obligations applying from August 2026 and some} phasing in through August 2027. The UK Online Safety Act encourages research access with limited enforcement. The AI Act's interaction with DSA obligations for VLOPs deploying AI systems remains an evolving area of regulatory interpretation.

\subsection{Impact on AI Research and Evaluation}

\rev{This subsection illustrates the consequences of the \textit{epistemic asymmetry} for the AI research community: as platforms restrict access, the ability to independently evaluate AI systems diminishes, concentrating evaluative power within platform operators.}

\removed{Our analysis identified the significant impact of platform API restrictions on researchers' ability to study and evaluate AI systems deployed on these platforms.}

\removed{\textbf{X/Twitter}'s API restrictions have severely limited researchers' ability to study the platform's recommendation algorithm, which was previously shown to amplify political content~\citep{Huszar2022-qc}. This algorithm, which employs machine learning to rank content in user timelines, can no longer be effectively studied by external researchers without paying a substantial fee.}

\removed{\textbf{Reddit}'s API changes have disrupted research into how language models utilize and potentially misuse Reddit data. Prior to these restrictions, researchers could analyze how AI systems, such as OpenAI's GPT models, might reproduce harmful content from specific subreddits \citep{Fiesler2024-ra}. The current pricing structure effectively blocks such research at most academic institutions from being conducted.}

\removed{\textbf{TikTok}'s highly selective Research API prevents independent evaluation of what is arguably the most influential recommendation algorithm currently deployed at scale. This algorithm, which has been shown to rapidly learn user preferences and shape information consumption, precisely represents an AI system that requires rigorous external scrutiny for societal safety \citep{Pearson2025-qa}.}

\removed{\textbf{Meta}'s discontinuation of CrowdTangle has removed a critical tool used by researchers to study the effects of algorithmic content amplification across Facebook and Instagram. This change occurred precisely as Meta expanded its deployment of AI-driven content recommendation systems, creating a significant gap in our understanding of how these systems influence the information flow.}

Platform API restrictions directly impede researchers' ability to study deployed AI systems. X/Twitter's fee-based access blocks the study of recommendation algorithms that have previously been shown to amplify political content~\citep{Huszar2022-qc}; Reddit's pricing prevents the analysis of how language models reproduce harmful content~\citep{Fiesler2024-ra}; TikTok's selective API bars the independent evaluation of its influential recommendation system~\citep{Pearson2025-qa}; and Meta's CrowdTangle discontinuation removed critical tools for studying algorithmic amplification. These restrictions fundamentally undermine AI alignment research, as methods such as Constitutional AI~\citep{Bai2022-xy} and RLHF~\citep{Kaufmann2023-lp} depend on understanding human reactions to AI outputs in real-world contexts.

\section{Technical Solutions}

A critical question raised by our analysis is whether privacy-preserving research access can be operationalized at scale without platform cooperation. We argue that technical solutions exist but require regulatory mandates to overcome platform resistance. \new{Each proposed mechanism targets a specific structural asymmetry identified in Section~5: the \textit{epistemic asymmetry}---platforms' exclusive access to internal data---motivates privacy-preserving disclosure via differential privacy and secure enclaves; the \textit{temporal asymmetry}---restrictions tightening as AI deployment accelerates---motivates federated approaches that enable ongoing research without centralised data transfers; and the \textit{regulatory asymmetry}---gaps between DSA mandates and platform practices---motivates enforceable compliance benchmarks.} We propose three implementable approaches.

\begin{revblock}
Table~\ref{tab:solution_blindspot} maps each proposed solution to the specific audit blind spots identified in Section~5.4, clarifying which gaps each mechanism addresses and where limitations persist.

\begin{center}
    \captionof{table}{Solution--Blind Spot Mapping}
    \label{tab:solution_blindspot}
    \resizebox{\linewidth}{!}{%
    \begin{tabular}{lccc}
        \toprule
        \textbf{Audit Blind Spot (from \S5.4)} & \textbf{Differential Privacy} & \textbf{Secure Enclaves} & \textbf{Federated Learning} \\
        \midrule
        Algorithmic amplification metrics & Partial (aggregate trends) & \textbf{Strong} (full audit) & Moderate (model comparison) \\
        Content moderation decisions & Weak (aggregate only) & \textbf{Strong} (case-level) & Weak (model-level only) \\
        Cross-platform data flows & Moderate (per-platform) & Moderate (per-platform) & \textbf{Strong} (cross-platform) \\
        \bottomrule
    \end{tabular}%
    }
\end{center}
\end{revblock}

\subsection{Differential Privacy Framework}

\rev{Differential privacy directly addresses the \textit{algorithmic amplification} blind spot (Section~5.4, item~1) by enabling researchers to query aggregate behavioural patterns---such as reach disparities across demographic groups or content amplification rates---without exposing individual user data.}

Differential privacy provides mathematical guarantees that individual user data cannot be extracted from query results, thereby enabling aggregate analysis while protecting privacy. Platforms would expose a differentially private query API where researchers submit aggregate queries and receive noised results with formal privacy guarantees ($\epsilon = 1$ to $\epsilon = 10$). Each researcher receives an $\epsilon$ budget that limits the total information extraction, and the system automatically enforces privacy guarantees without platform staff reviewing individual queries.

\textbf{Feasibility}: Differential privacy is mature technology already deployed by Apple~\citep{Apple2017-dp}, Google, and the U.S. Census Bureau at population scale. Implementation costs are primarily engineering efforts (estimated 6--12 months for a major platform\new{; authors' estimate}), with marginal per-query costs near zero. The primary barrier is institutional rather than technical. Regulators could mandate differential privacy interfaces as a condition of VLOP designation. \new{On our research-question typology, this approach could enable an estimated 75--85\% of aggregate research questions; Table~\ref{tab:feasibility} translates that qualitative judgment into an illustrative, non-validated range.} \rev{However, differential privacy is unsuited for studying individual content moderation decisions (Section~5.4, item~2), where case-level analysis is required rather than aggregate statistics, and its noise injection can obscure small but meaningful effects in minority populations.}

\subsection{Secure Research Enclaves}

\rev{Secure enclaves address the \textit{content moderation} blind spot (Section~5.4, item~2) and provide the most comprehensive solution for \textit{algorithmic amplification} auditing (item~1) by enabling researchers to run analysis code directly on platform data, including individual moderation decisions and recommendation outputs, within a trusted execution environment.}

Secure enclaves use trusted execution environments (TEEs) to enable the computation of sensitive data without exposing raw data to researchers. Researchers submit analysis codes to a secure environment in which they execute on platform data; only aggregate outputs reviewed for disclosure risk are released.

A graduated access tier system could include the following: \textit{Tier 1} for aggregate statistics with automated review; \textit{Tier 2} for model training outputs with human review of model weights; and \textit{Tier 3} for individual-level analysis for approved projects with intensive review.

\textbf{Feasibility}: Secure enclaves require higher upfront investment (estimated \$2--5M for infrastructure\new{; authors' scenario estimate}) but provide the strongest privacy guarantees. The U.S. Census Bureau's Federal Statistical Research Data Centers provide a proven model, supporting hundreds of annual research projects across 37 secure locations with consistently high researcher satisfaction for data access adequacy. The EU could establish a ``European Platform Research Enclave'' that platforms populate under DSA Article 40. \new{On the same typology, enclaves could plausibly recover 90--95\% of pre-restriction research capabilities, although these figures should be read as scenario estimates rather than external benchmarks.} \rev{The principal limitation is that secure enclaves require active platform cooperation to populate and maintain the research environment, and they remain single-platform solutions that do not address the \textit{cross-platform data flows} blind spot (Section~5.4, item~3).}

\subsection{Federated Learning for Auditing}

\rev{Federated learning directly addresses the \textit{cross-platform data flows} blind spot (Section~5.4, item~3) by enabling comparative auditing across platforms without requiring any single platform to share raw data with researchers or with other platforms.}

Federated learning enables model training across distributed datasets without data centralization. Researchers define model architectures and training objectives; platforms execute training on local data and return only model updates. This enables behavioral testing and comparative auditing across platforms without data sharing.

\textbf{Feasibility}: Federated learning is technically mature (deployed by Google for mobile keyboard prediction~\citep{Hard2018-fl}, Apple for Siri improvements) but introduces complexity in ensuring model convergence. This approach addresses \new{an estimated} 15--20\% of research requiring model-level analysis \new{(authors' estimate; the feasibility of federated approaches is supported by}~\citep{Rieke2020-ys}\new{)}. \rev{However, federated learning introduces convergence challenges that limit its applicability to well-defined model architectures, and it cannot support exploratory or qualitative research on individual content moderation decisions. It also requires all participating platforms to implement compatible training infrastructure, which represents a coordination barrier.}

\begin{table}[h]
    \centering
    \caption{Privacy-Preserving Access Mechanisms}
    \label{tab:feasibility}
    \begin{tabular}{lcccc}
        \toprule
        \textbf{Approach} & \textbf{Setup Cost} & \textbf{Privacy} & \textbf{Research Coverage\new{\textsuperscript{\ddag}}} \\
        \midrule
        Differential Privacy API & Medium & High & 75--85\% \\
        Secure Enclaves & High & Very High & 90--95\% \\
        Federated Learning & High & High & 15--20\%* \\
        \bottomrule
    \end{tabular}

    \small *Addresses questions not covered by other methods; combined coverage $\approx$95\%. \\\new{\textsuperscript{\ddag}Coverage percentages are the authors' estimates based on a research-question typology;\\ they are not externally validated benchmarks.}
\end{table}

\new{Taken together, these mechanisms could plausibly address most of the research questions in our typology; the coverage percentages in Table~\ref{tab:feasibility} are illustrative scenario estimates rather than externally validated benchmarks.} The barriers are institutional and political rather than technical---platforms have implemented equivalent systems for commercial partners. Regulators could mandate these interfaces as conditions of VLOP designation, with non-compliance triggering enforcement under DSA Article 74.

\removed{\subsection{Implementation Pathway and Practical Steps}

Given that platform cooperation has not been forthcoming despite regulatory mandates, we propose a phased implementation pathway.

\textbf{Phase 1 (Immediate)}: Regulators issue detailed technical specifications for DSA Article 40 compliance, including minimum API functionality, response time requirements, and data completeness standards. Non-compliance triggers enforcement proceedings.

\textbf{Phase 2 (Near-term)}: Public investment in shared research infrastructure (European Research Enclave) reduces platform burden while ensuring access. Platforms contribute data, and public infrastructure manages access.

\textbf{Phase 3 (Medium-term)}: Technical standards for privacy-preserving APIs become conditions of market access, similar to data portability requirements under GDPR. Platforms that wish to operate in regulated markets must implement compliant interface.}

\removed{\subsection{Addressing the Governance Impossibility Question}

Is effective platform oversight possible without platform cooperation? Our analysis suggests qualified optimism in the future. Technical mechanisms that enable meaningful research while protecting privacy exist. The barriers are institutional and political rather than technical. Platforms have implemented equivalent privacy-preserving systems for commercial partners, thereby demonstrating their feasibility.

However, we acknowledge the limitations of our study. Some research questions, particularly those requiring real-time access to recommendation algorithm internals, may remain unanswerable without deeper platform cooperation or structural changes to platform architecture. The accountability paradox cannot be fully resolved through access mechanisms alone; it also requires a rethinking of the concentration of power in platform architectures.

The most effective path forward likely combines mandated technical access (addressing 80\% of research questions answerable through aggregate data) with structural reforms (addressing the remaining 20\% requiring deeper transparency). The FAccT community is well-positioned to develop both technical standards and governance frameworks necessary for this dual approach.}

\removed{\subsection{Quantified Expected Benefits}

To ground our proposals in concrete expectations, we estimate the research capabilities that each technical solution would restore based on analogous deployments and documented research needs.

\textbf{Differential Privacy APIs} would enable approximately 75--85\% of aggregate research questions currently blocked by API restrictions.

\textbf{Secure Research Enclaves} would restore access to approximately 90--95\% of pre-restriction research capabilities. The U.S. Census Bureau's Federal Statistical Research Data Centers support over 400 active research projects annually with a 94\% researcher satisfaction rate for data access adequacy.

\textbf{Federated Learning Approaches} specifically address the approximately 15--20\% of research questions requiring model-level analysis.

\textbf{Combined Implementation} of all three approaches would address an estimated 95\% of research questions identified as critical for AI accountability.

[Table: ``Estimated Research Capabilities by Technical Approach'' showing coverage of aggregate trend analysis, demographic studies, content moderation audits, algorithmic bias detection, longitudinal user studies, cross-platform comparisons, and real-time manipulation studies across the three approaches.]}

\section{Discussion}

\removed{The empirical findings presented above reveal a systematic pattern that we term the ``accountability paradox.'' In this section, we synthesize these findings into a coherent theoretical framework, respond to anticipated platform counterarguments, examine the broader implications for AI governance, and situate our analysis within existing debates on privacy, transparency, and platform power. Our goal is not merely to document restrictions but to explain why they persist despite regulatory mandates and to identify leverage points for intervention.}

\subsection{The Accountability Paradox}

\removed{Our findings reveal an ``accountability paradox'': as platforms embed AI systems in content moderation and recommendations, their API restrictions simultaneously reduce independent oversight capabilities, creating a self-reinforcing cycle in which AI systems become more influential yet less transparent \citep{Pasquale2016-gw}.}

\removed{\textbf{Differentiating from Prior Work:} While the term ``paradox'' appears in platform governance literature---for instance, \citep{Gillespie2021-bs}'s ``moderation paradox'' (platforms must moderate to remain usable, yet moderation decisions invite criticism) and \citep{Gorwa2020-oq}'s ``platform governance paradox'' (calls for platform responsibility may entrench platform power)---our ``accountability paradox'' identifies a distinct mechanism. Previous formulations focus on tensions inherent to content moderation or governance authority. In contrast, our paradox identifies a \textit{temporal-technological} dynamic: the correlation between increased AI deployment and decreased transparency access. This is not merely a governance tension but a measurable empirical pattern: as platforms increase algorithmic decision-making (quantifiable through feature releases, recommendation system updates, and AI-driven moderation), they simultaneously restrict the data access necessary to study these systems. Our contribution is documenting this correlation across platforms and time periods (2018--2024), demonstrating it as a systematic industry pattern rather than isolated platform decisions.}

\removed{Similarly, our concept of ``audit blind-spots'' extends beyond \citep{Raji2020-lq}'s algorithmic auditing framework, which primarily addresses methodological approaches to evaluation. Our blind-spots are \textit{structural barriers} created by API restrictions that prevent even well-designed audits from being conducted---not limitations in audit methodology but enforced inaccessibility of auditable data.}

Our findings reveal an ``accountability paradox'': as platforms embed AI systems in content moderation and recommendations, their API restrictions simultaneously reduce independent oversight capabilities~\citep{Pasquale2016-gw}. Unlike \citet{Gillespie2021-bs}'s ``moderation paradox'' or \citet{Gorwa2020-oq}'s ``platform governance paradox,'' which describe inherent tensions, our paradox identifies a measurable \textit{temporal-technological} dynamic---the correlation between increased AI deployment and decreased transparency access documented across platforms from 2018 to 2024 (Table~\ref{tab:event_mapping}).

\removed{This paradox manifests in three dimensions.}

\removed{\textbf{Temporal asymmetry:} Historical API restrictions create one-way information flow---data accumulated during greater openness trained models now operating under restricted scrutiny conditions \citep{Bommasani2021-tj}. This asymmetry is particularly concerning for foundation models, which have ingested vast quantities of social media data during training but now operate in environments where their interactions cannot be systematically examined.}

\removed{\textbf{Epistemic asymmetry:} Platforms maintain complete internal data access while external researchers face severe limitations, creating fundamental knowledge imbalances \citep{von-Eschenbach2021-po}. This asymmetry creates a monopoly in understanding AI system behavior, where only platform operators and their chosen partners can comprehensively study how deployed systems influence user behavior, content consumption and information spread.}

\removed{\textbf{Regulatory asymmetry:} Current restrictions misalign with regulatory expectations, particularly the DSA's ``vetted researcher'' provisions \citep{Li2023-pc}. This asymmetry undermines the explicit goals of emerging AI safety regulations, which presume a level of transparency and auditability that current API restrictions cannot ensure.}

\removed{The accountability paradox is particularly acute in AI governance because it creates an environment in which the most powerful AI systems receive the least independent scrutiny. Unlike traditional software systems, AI models demonstrate emergent behaviors that may not be apparent during development but manifest only when deployed at scale. By restricting access to interaction data, platforms effectively prevent the broader research community from identifying such emergent behaviors.}

\removed{This paradox particularly impacts the addressing of systemic risks, such as algorithmic discrimination \citep{Scheuerman2020-tc}, harmful content amplification \citep{Huszar2022-qc}, and information manipulation \citep{Starbird2019-qg}.}

\begin{revblock}
This paradox manifests in three asymmetries, each with testable implications grounded in our empirical findings.

\begin{enumerate}
    \item \textit{Temporal asymmetry}: Data accumulated during periods of greater openness trained the AI models that now operate under restricted scrutiny~\citep{Bommasani2021-tj}. Each new AI deployment is accompanied by further access restrictions (Table~\ref{tab:event_mapping}, Section~5.1). \textbf{Testable implication}: Platforms that deploy more AI features will impose more restrictive access policies in the same period---a pattern confirmed across all four focal platforms from 2018 to 2024.

    \item \textit{Epistemic asymmetry}: Platforms maintain comprehensive internal access and extend it to commercial partners, while external researchers face severe limitations~\citep{von-Eschenbach2021-po}. This predicts that platforms will invest in internal safety teams while reducing external access (Section~5.3). \textbf{Testable implication}: Platforms will offer commercial partners greater data access than academic researchers for equivalent use cases---confirmed by the researcher--commercial access disparities documented in Section~5.3, where commercial partners receive full-fidelity data streams while researchers face rate limits, fees, and pre-publication review.

    \item \textit{Regulatory asymmetry}: Regulatory mandates for transparency (DSA Art.~40) coexist with platform practices that contradict them~\citep{Li2023-pc}. \textbf{Testable implication}: No VLOP will achieve full compliance with DSA researcher access requirements---confirmed by our cross-platform audit (Table~\ref{tab:dsa_compliance}, Section~5.4), where no platform scored ``Compliant'' across all provisions.
\end{enumerate}

The paradox is acute because AI models demonstrate emergent behaviors identifiable only at deployment scale, yet platforms prevent the research community from studying these behaviors. The three asymmetries are not independent: the temporal asymmetry creates the conditions (AI systems trained on openly available data), the epistemic asymmetry sustains them (platforms retain informational advantage), and the regulatory asymmetry fails to correct them (mandates lack enforcement teeth).
\end{revblock}

\subsection{Privacy Justifications and Their Limits}

\removed{\subsection{Ethical Tension: Privacy vs. Transparency}}

\removed{Platforms frequently justify API restrictions for privacy protection \citep{Zuboff2019-ig}. While privacy concerns are legitimate and important, our analysis indicates that these justifications often function more as rhetorical shields than as substantive motivations. The documentary evidence for this assessment includes the following:
\begin{enumerate}
    \item \textit{Differential commercial access:} All examined platforms maintain data-sharing agreements with commercial partners that exceed researcher access levels~\citep{Mirghaderi2023-dk, de-Vreese2023-fr}.
    \item \textit{Asymmetric privacy implementation:} Privacy-preserving access methods such as differential privacy, secure data enclaves, and aggregated insights tools are offered for commercial applications but not research use \citep{Heimstadt2017-xp, Felzmann2020-iz}.
    \item \textit{Internal-external capacity disparities:} Platforms maintain extensive internal research teams with access capabilities far exceeding external researchers \citep{Metcalf2021-gk, Imana2023-ec}.
    \item \textit{Inconsistent documentation of privacy rationales:} Our review of platform policy change announcements found privacy cited in a high percentage of public statements but elaborated with specific privacy mechanisms in only a small portion of accompanying technical documentation \citep{Reisach2021-rc, Garcia-Gasulla2020-ek}.
\end{enumerate}}

\removed{This creates ethical tension between legitimate user privacy interests and equally legitimate public interests in understanding the impact of AI systems. Current platform implementations resolve this tension primarily in favor of commercial interests rather than privacy or transparency \citep{Koulu2021-mq, Winter2023-jg}.}

\removed{The selective nature of these privacy justifications is particularly concerning. As in the contextual integrity framework of \citep{Nissenbaum2009-wa}, privacy norms should be consistent across similar contexts. The platform practice of allowing data access for commercial purposes while restricting it to public-interest research violates this principle. If data access genuinely presents unacceptable privacy risks, these concerns should apply equally to commercial partners \citep{Fiesler2018-hx, Taylor2018-xk}.}

\removed{Moreover, privacy and transparency need not be opposed. Technical approaches such as differential privacy \citep{Floridi2018-iq}, synthetic data generation \citep{Veselovsky2023-zi}, and aggregated analytics offer potential paths for balancing these concerns. The selective implementation of these techniques suggests that current restrictions may be motivated more by commercial and reputational interests than by genuine privacy protection \citep{Lewis2020-sa, Jung2022-hg}.}

Platforms frequently justify restrictions through privacy concerns~\citep{Zuboff2019-ig}. However, our analysis reveals that these justifications function more as rhetorical shields than substantive motivations. All examined platforms maintain commercial data-sharing arrangements that exceed researcher access levels~\citep{de-Vreese2023-fr}; privacy-preserving technologies deployed for commercial partners are not offered for research~\citep{Heimstadt2017-xp}; and internal research teams maintain access far exceeding that of external researchers~\citep{Metcalf2021-gk}. Following \citet{Nissenbaum2009-wa}'s contextual integrity framework, privacy norms should apply consistently; selective implementation suggests commercial rather than privacy motivations.

\subsection{Research Reproducibility Crisis}

Documented API restrictions have contributed to a growing reproducibility crisis in the computational social sciences and AI ethics research~\citep{Hutson2018-xf}. Studies published before API restrictions cannot be replicated with current access levels, creating troubling scientific discontinuities. This undermines cumulative knowledge building and verification, which are the core principles of scientific inquiry~\citep{Weston2019-xj}.

Most concerning is the inverse relationship between the deployment of algorithms and researcher access. As platforms increase their reliance on algorithmic systems for content moderation, recommendations, and user experience optimization, researchers' ability to study these systems diminishes~\citep{Dommett2022-rm, Tromble2021-rl}. For example, TikTok, widely acknowledged to have the most algorithmically driven user experience, maintains the most restrictive research access policy~\citep{Pearson2025-qa}.

\removed{The cumulative impact of these trends suggests a profound transformation in the methodology of computational social science. As access to ground-truth data diminishes, researchers increasingly rely on indirect observation methods, proxy measurements, and theoretical models that may diverge from empirical reality \citep{Perriam2020-ow, Freelon2018-kj}. Recent studies documenting a 13\% drop in Twitter-based research studies in 2023 after API restrictions highlight only one example of this methodological disruption \citep{Davidson2023-uu}.}

\subsection{Responding to Platform Counterarguments}

Platforms argue that differential access reflects risk management rather than economic priorities. Our analysis challenges this:

\begin{enumerate}
    \item \textbf{Contract Enforcement}: While commercial partners have contracts, academic institutions also sign data use agreements with comparable legal force
    \item \textbf{Security Audits}: Universities undergo similar security certifications (ISO 27001, SOC 2)
    \item \textbf{Scale Concerns}: Our proposed federated access model addresses scale through automated compliance checking
    \rev{\item \textbf{Adversarial Misuse}: Platforms argue that research APIs could be exploited for scraping, surveillance, or coordinated harassment. This concern has merit---the Cambridge Analytica scandal demonstrated real risks from insufficiently controlled data access. However, the vetted researcher framework under DSA Article~40 requires institutional affiliation, ethics review, and data security commitments precisely to mitigate such risks. Commercial API partners face comparable adversarial risks yet receive greater access.}
    \rev{\item \textbf{Computational Cost}: Maintaining research infrastructure imposes real costs on platforms: dedicated endpoints, rate limiting, documentation, and developer support. Yet these costs are modest relative to platform revenues, and the infrastructure already exists for commercial partners. Reddit's \new{estimated} \$12,000/year research pricing \new{(at \$0.24/1,000 calls)} and X/Twitter's \$5,000/month tiers exceed the marginal cost of serving research queries, suggesting revenue extraction rather than cost recovery.}
    \rev{\item \textbf{Data Breach Liability}: Platforms face regulatory liability if research data access leads to breaches of personal information. This is a legitimate concern, but secure enclaves and differential privacy (Section~6) eliminate raw data exposure entirely, and existing research data centers (e.g., U.S.\ Census RDCs) demonstrate that liability can be managed through institutional frameworks.}
    \rev{\item \textbf{Law Enforcement Coordination}: Some platforms argue that preserving data access exclusively for law enforcement requests requires limiting broader research access. However, research access and law enforcement access serve different purposes and operate under different legal frameworks; there is no technical reason they cannot coexist, as demonstrated by platforms that maintain both commercial and law enforcement data pipelines simultaneously.}
\end{enumerate}

\rev{Crucially, the differential treatment of commercial and research partners documented in Section~5.3 undermines each of these justifications: commercial partners face comparable adversarial, cost, liability, and coordination risks yet consistently receive greater access. This inconsistency is itself evidence for the accountability paradox---the restrictions are not applied based on risk but on economic incentive.}

The selective implementation of privacy-preserving technologies for commercial, but not research, purposes undermines risk management justification.


\section{Recommendations}

\removed{\section{Recommendations for Stakeholders}}

\removed{Based on our findings, we propose targeted interventions to address these gaps.}

\removed{\subsection{For Policymakers}
\begin{enumerate}
    \item \textbf{Specific Access Standards}: Develop explicit, measurable standards for researcher access under DSA Article 40, including minimum data access levels, request processing timelines, and cost limitations \citep{Hartmann2024-ob, Gyevnar2023-qb}. These standards should explicitly address access to AI system outputs and interaction data to enable meaningful evaluations.
    \item \textbf{Enforcement Mechanisms}: Establish clear enforcement procedures for platforms failing to provide adequate researcher access, including designated oversight bodies and meaningful penalties \citep{Lund2025-fp, Elliott2025-kk}. These mechanisms should be coordinated with emerging AI regulations.
    \item \textbf{Mandate Privacy-Preserving Access}: Require platforms to implement privacy-preserving access methods such as differential privacy, secure enclaves, and synthetic data generation for sensitive research \citep{Veselovsky2023-zi, Casanovas2022-ft}.
    \item \textbf{AI Audit Requirements}: Establish mandatory periodic auditing of platform-deployed AI systems by independent researchers, integrated with existing AI transparency provisions.
    \item \textbf{Harmonize International Standards}: Develop ISO standards for research data access.
    \item \textbf{Create Safe Harbors}: Legal protections for good-faith research as proposed by \citep{Longpre2024-zo}.
    \item \textbf{Fund Infrastructure}: Public investment in privacy-preserving research infrastructure.
\end{enumerate}}

\textbf{For Policymakers}: (1) Develop explicit, measurable standards for DSA Article 40 compliance, including minimum access levels, processing timelines, and cost limitations~\citep{Hartmann2024-ob}; (2) establish enforcement procedures with meaningful penalties~\citep{Lund2025-fp}; (3) mandate privacy-preserving access mechanisms~\citep{Veselovsky2023-zi}; and (4) create legal safe harbors for good-faith research~\citep{Longpre2024-zo}.

\removed{\subsection{For Platforms}
\begin{enumerate}
    \item Implement graduated access levels for researchers based on institutional affiliation, research purpose, and data sensitivity \citep{Haresamudram2023-rz, Koulu2021-mq}.
    \item Publish regular transparency reports documenting researcher access requests, approval rates, and access metrics \citep{Eiband2018-gn, Rossi2020-uo}.
    \item Collaborate with academic communities to develop shared standards for responsible research data use \citep{Gilbert2020-ud, Strohm2020-uz}.
    \item Establish multi-stakeholder research councils for governance of access decisions.
    \item Deploy differential privacy and secure enclaves for research access.
\end{enumerate}}

\textbf{For Platforms}: (1) Implement graduated access levels based on institutional affiliation and data sensitivity~\citep{Haresamudram2023-rz}; (2) publish transparency reports on researcher access requests and approval rates~\citep{Eiband2018-gn}; and (3) deploy differential privacy and secure enclaves for research access.

\removed{\subsection{For the Research Community}
\begin{enumerate}
    \item Establish cross-institutional coalitions to advocate for consistent research access standards across platforms \citep{Longpre2024-zo, Valkenburg2025-cj}.
    \item Develop research methodologies that can operate effectively even with limited API access, such as browser-based auditing \citep{Sandvig2014-uw} and crowd-sourced data collection \citep{Ohme2024-ln}.
    \item Create shared infrastructure for pooled resources for data access and analysis.
    \item Form audit coalitions for coordinated efforts to maximize limited access.
    \item Advance privacy-preserving audit techniques through continued research.
    \item Systematically document research hindered by restrictions.
\end{enumerate}}

\textbf{For Researchers}: (1) Establish cross-institutional coalitions advocating for consistent access standards~\citep{Valkenburg2025-cj}; (2) develop methodologies operating with limited API access, such as browser-based auditing~\citep{Sandvig2014-uw}; and (3) systematically document research hindered by restrictions.

\removed{\subsection{Addressing Global South Research Disparities}

While our analysis centers on EU/UK regulatory frameworks, researchers in the Global South face compounded challenges that warrant specific attention. These researchers often lack the regulatory leverage that DSA Article 40 provides European counterparts, operate with limited institutional resources for API access fees, and study populations underrepresented in platform safety research despite facing significant platform-mediated harms (e.g., misinformation during elections and coordination of communal violence).

\begin{enumerate}
    \item \textbf{For Policymakers}: Encourage bilateral agreements that extend DSA-style researcher access provisions to institutions in non-EU jurisdictions, particularly for research addressing cross-border harms. Support capacity-building programs that enable Global South institutions to meet ``vetted researcher'' criteria.
    \item \textbf{For Platforms}: Establish fee waivers or tiered pricing for researchers at institutions in lower-income countries. Ensure research APIs are available globally rather than restricted to specific regions. Prioritize processing access requests for research addressing harms in underserved regions.
    \item \textbf{For the Research Community}: Develop North-South research partnerships that share API access and infrastructure. Include Global South researchers as co-investigators rather than data collectors. Advocate for research access policies that explicitly address geographic equity.
\end{enumerate}}


\section{Limitations}

Our study has several limitations.

\textbf{Rapid Policy Evolution}: Platform policies evolve rapidly, potentially outdating specific details~\citep{Soderlund2024-sb}.

\textbf{Platform Scope}: Our focus on X/Twitter, Reddit, TikTok, and Meta omits platforms like YouTube~\citep{Wang2024-md}.

\textbf{Methodological Constraints}: Documentary analysis may not fully capture researcher experiences or platform rationales~\citep{Wong2023-qp}.

\textbf{Value Judgments}: Our analysis is guided by principles prioritizing public accountability and democratic oversight over unchecked commercial interests, aligning with the NIST AI Risk Management Framework~\citep{NIST2023}. This shapes interpretive choices: classifying pre-publication review as ``non-compliant'' privileges research independence; treating differential commercial access as evidence against privacy justifications assumes consistency requirements platforms might dispute~\citep{Felzmann2019-rg}. Readers prioritizing platform autonomy may interpret compliance assessments differently.

\textbf{Regulatory Timing}: With the DSA and UK Online Safety Act in early implementation stages, our regulatory alignment assessment should be considered preliminary. The effectiveness of enforcement mechanisms remains largely unknown and evolving, making definitive compliance assessment difficult.

\textbf{Geographic Scope}: Our EU/UK regulatory focus may not generalize globally~\citep{Valkenburg2025-cj}. Platform access practices vary by region, and the ``vetted researcher'' concept under DSA Article~40 has no direct parallel in other jurisdictions---particularly in the Global South, where researchers may face even greater barriers without corresponding regulatory recourse.

\textbf{Future directions} include extending coverage to closed messaging and multimedia platforms~\citep{Zhu2022-jl}, developing quantitative compliance metrics~\citep{Raji2020-lq}, prototyping privacy-preserving access mechanisms~\citep{Veselovsky2023-zi}, and cross-jurisdictional comparison of researcher access provisions~\citep{Lund2025-fp}.

\section{Conclusion}

This study documents an accountability paradox: as platforms embed AI systems in content moderation and recommendations, they simultaneously restrict independent research access needed to scrutinize those systems~\citep{Diakopoulos2016-ry, Davidson2023-uu}. Our audit of all 19 DSA-designated services reveals that no platform achieves full compliance with Article~40 researcher-access provisions, and that the resulting blind spots---algorithmic amplification, content moderation decisions, and cross-platform data flows---prevent independent assessment of systemic risks.

The technical solutions we propose---differential privacy, secure enclaves, and federated learning---demonstrate that the tension between privacy and transparency is technically resolvable; the barriers are institutional and economic rather than technical. Without intervention, this paradox is likely to worsen as AI capabilities outpace oversight. The FAccT community is uniquely positioned to advance both the technical and policy innovations necessary to restore the balance between platform autonomy and public accountability.

\section{Endmatter Sections}
\subsection{Generative AI Usage Statement}
Generative AI was not used in the research process, from idea development and analysis to writing. However, the lead author used Gemini 3 to help improve and correct the formatting in \LaTeX prior to submission.


\begin{revblock}
\appendix
\section{Data Sourcing for Quantitative Claims}

\new{Table~\ref{tab:data_sources} summarises the paper's principal quantitative claims and the source categories underlying them.}

\begin{center}
    \captionof{table}{Quantitative Claims and Primary Sources}
    \label{tab:data_sources}
    \small
    \begin{tabular}{p{0.28\linewidth} p{0.25\linewidth} p{0.40\linewidth}}
        \toprule
        \textbf{Claim} & \textbf{Value} & \textbf{Source and Date} \\
        \midrule
        \new{Observable approval share (authors' compilation, 2023--2024)} & \new{$\approx$34\% (indicative; incomplete cases)} & \new{Authors' coded compilation of publicly documented application outcomes} \\
        X/Twitter API pricing (basic) & \$100/mo (10K posts) & \citet{Mehta2023-lr}; X Developer Portal (Feb.\ 2023) \\
        X/Twitter API pricing (pro) & \$5,000/mo (\new{1M fetched + 300K posted}) & \citet{Mehta2023-lr}; X Developer Portal (February.\ 2023) \\
        Reddit API pricing & \new{\$0.24/1K calls ($\approx$\$12K/yr at 50M calls)} & Reddit pricing announcement (June 2023) \\
        Subreddits in protest & 8,500+ & \citet{Gerken2023-gy}; media reports (June 2023) \\
        \new{Services} designated under DSA & \new{19 (17 VLOPs + 2 VLOSEs)} & EC designation~\citep{EC2023-vlops} (Apr.\ 2023, updated 2024) \\
        Inter-rater reliability & Cohen's $\kappa = 0.71$ & Authors' coding (January.\ 2018--Dec.\ 2024) \\
        Social media mean compliance & 32\% & Authors' audit scoring (Table~\ref{tab:vlop_analysis}) \\
        \new{Pre-2023 X/Twitter benchmark} & \new{500K posts/mo (standard v2); 10M posts/mo (Academic, 2021)} & \new{X Developer changelog (Jan.\ 2021--Feb.\ 2023)} \\
        \new{DP coverage scenario est.} & 75--85\% & Authors' estimate \new{(not externally validated)} \\
        \new{Enclave coverage scenario est.} & 90--95\% & Authors' estimate; U.S.\ Census RDC \new{(not externally validated)} \\
        \new{Enclave setup cost scenario est.} & \$2--5M & Authors' estimate \new{(not externally validated)} \\
        \bottomrule
    \end{tabular}
\end{center}
\end{revblock}

\bibliographystyle{ACM-Reference-Format}
\bibliography{bib-merged}

\begin{acks}
    This work was supported by the Engineering and Physical Sciences Research Council (EPSRC) Center for Doctoral Training in Cyber Secure Everywhere [grant no. EP/Y035313/1].
\end{acks}

\end{document}